\def\cm2{cm$^{-2}$}

\def\c2{C~{\sc ii}}
\def\c4{C~{\sc iv}}
\def\fe2{Fe~{\sc ii}}
\def\fe3{Fe~{\sc iii}}
\def\mg1{Mg~{\sc i}}
\def\mg2{Mg~{\sc ii}}
\def\si2{Si~{\sc ii}}
\def\si4{Si~{\sc iv}}
\def\al2{Al~{\sc ii}}
\def\al3{Al~{\sc iii}}
\def\o1{O~{\sc i}}
\def\n1{N~{\sc i}}
\def\h1{H~{\sc i}}

\def\approxlt{\mathrel{\spose{\lower 3pt\hbox{$\sim$}}
        \raise 2.0pt\hbox{$<$}}}
\def\approxgt{\mathrel{\spose{\lower 3pt\hbox{$\sim$}}
        \raise 2.0pt\hbox{$>$}}}

\documentclass[article]{gwp80}
\usepackage{graphicx}
\usepackage{amssymb}
\tabletypesize{\normalsize}  

\shortauthors{Amigo et al.}
\shorttitle{New Variable Stars in CMa Globular Cluster Candidates}

\begin{document}
\large    
\pagenumbering{arabic}
\setcounter{page}{101}

\title{New Variable Stars in CMa Globular Cluster Candidates}

%
%
\author{{\noindent P. Amigo{$^{\rm 1\dag}$}, M. Catelan{$^{\rm 1}$}, P. B. Stetson{$^{\rm 2}$}, H. A. Smith{$^{\rm 3}$}, C. Cacciari{$^{\rm 4}$}, M. Zoccali{$^{\rm 1}$}\\
\\
{\it (1) Departamento de Astronomia y Astrofisica,  Pontificia Universidad Cat\'olica de Chile, Santiago, Chile\\
(2) Dominion Astrophysical Observatory, Herzberg Institute of Astrophysics, National Research Council of Canada, Victoria, BC, Canada\\
(3) Department of Physics and Astronomy, Michigan State University, East Lansing, MI, USA\\
(4) INAF-Bologna, Italy} 
}
}

%
%
\email{(\dag) pamigo@astro.puc.cl }

\begin{abstract}

We present the preliminary results of an image-subtraction analysis of the Galactic globular cluster M79 (NGC 1904), 
as well as a new investigation of the variable star population in NGC~1851. Both M79 and NGC~1851 
have been previously associated with the Canis Major overdensity, which has been suggested to have an extragalactic origin.
We found 6 new RR Lyrae in M79, and also recovered 3 previously known RR Lyrae. The average period of the 5 ab-type RR Lyrae 
is 0.68~d, corresponding to an Oosterhoff II classification~-- which is unusual, though not unprecedented, for systems of 
extragalactic origin. We also report on the discovery of at least 4 previously unknown variables in NGC~1851. 
\end{abstract}

\section{Introduction}
The leading scenario for the formation of the Galaxy envisages the accretion of smaller galaxies, 
disrupted and accreted over time (Searle \& Zinn 1978). This originally largely qualitative scenario has received
considerable support from simulations carried out within the framework of the $\Lambda$CDM cosmological paradigm, 
which clearly predict that large galaxies such as the Milky Way are formed in a hierarchical fashion \citep[e.g.,][]{jdea07}.
In this scenario, dense, massive globular clusters could survive the accretion proccess (e.g., Forbes \& Bridges 2010). 
The existence of the Sagittarius dwarf spheroidal galaxy and stream (Ibata et al. 1994), which is clear evidence of ongoing accretion, 
gives support to this scenario. 

In this sense, another possible case of a dwarf galaxy being accreted by the Milky Way is provided by the Canis Major overdensity
(Martin et al. 2004), which may be related to the Monoceros stream. In this case, it has been matter of considerable debate whether 
the CMa system represents a true extragalactic system or  rather an extension of the warp of the outer Galactic disk. 
(see Sbordone et al, 2005; Momany et al., 2004; Bellazzini et al., 2004; Carraro et al., 2008 for more discussion on this topic)

One way to shed light on the possibility of an old star cluster being associated with an accreted dwarf galaxy may be provided
by the so-called Oosterhoff (1939) dichotomy. This is a very interesting property of the RR Lyrae population in Galactic globular clusters,
which divide into two classes according to their ab-type RR Lyrae pulsation properties: Oosterhoff type I (OoI),  with $\langle P_{\rm ab} \rangle \sim 0.55$~d, 
and Oosterhoff type II (OoII), with $\langle P_{ab} \rangle \sim 0.65$~d. The period range between 0.58~d and 0.62~d is poorly populated,
and has been called the ``Oosterhoff gap'' (Catelan 2004). Field stars in the Milky Way halo also seem to display the Oosterhoff dichotomy
(e.g., Miceli et al. 2008; Szczygiel et al. 2009). 
Intriguingly, the Oosterhoff dichotomy is observed only in the Milky Way: nearby extragalactic systems show instead a single-peaked
$\langle P_{\rm ab} \rangle$ distribution, with the peak located inside the Oosterhoff gap. Thus, extragalactic systems are predominantly
{\em Oosterhoff-intermediate}, and therefore an Oosterhoff-intermediate classification for a globular cluster may provide a hint of a possible
extragalactic origin (see Catelan 2006, 2009 for recent reviews and additional discussion).  
\begin{flushleft}
\begin{deluxetable*}{ccccl}[h!]
\tabletypesize{\normalsize}
\tablecaption{Positions and periods for RR Lyrae in M79}
\tablewidth{0pt}
\tablehead{ \\ \colhead{Variable ID}   & \colhead{RA (J2000)} &
       \colhead{DEC (J2000)} & \colhead{Period (days)} & \colhead{Type} \\
}

\startdata
 V4\rlap{\tablenotemark{b}} & 05 24 17.76   & -24 32 16.3  &  0.6341531 & RRab    \\
 V3\rlap{\tablenotemark{b}} & 05 24 13.52   & -24 32 29.2  &  0.7350907 & RRab    \\
 V6\rlap{\tablenotemark{b}} & 05 24 06.01   & -24 29 32.9 &  0.3387880 & RRc     \\
 NV1 &05 24 12.56 & -24 31 52.8 &  0.3616000 & RRc     \\
 NV2 &05 24 12.10 & -24 31 34.2 &  0.7279145 & RRab    \\
 NV3 &05 24 11.92 & -24 31 34.5 &  0.8199846 & RRab??  \\
 NV4 &05 24 11.37 & -24 31 28.4 &  0.3234196 & RRc     \\
 NV5 &05 24 10.57 & -24 31 11.5 &  0.6906617 & RRab    \\
 NV6 &05 24 10.22 & -24 31 03.7 &  0.6683208 & RRab    \\

 \enddata                                    
\tablenotetext{a}{ID of previously known variables, according to Clement's catalogue}
\label{TaM79}
\end{deluxetable*}
\end{flushleft}

In this study we analyze time-series photometric data for the globular cluster (M79), which has been associated with the Canis Major overdensity
(e.g., Forbes et al. 2004). In particular, we show the preliminary results of a new search for variable stars, and the periods obtained for the RR Lyrae 
stars found. We also present new time-series photometry for NGC~1851, also associated with Canis Major, including the discovery
of 10 previously unknown variables. The Oosterhoff types of the globular clusters found so far in CMa are then discussed within the framework 
of the potential extragalactic origin of this system.

\section{Variable stars in M79}

   \begin{figure*}
   \centering
  \includegraphics[width=15cm]{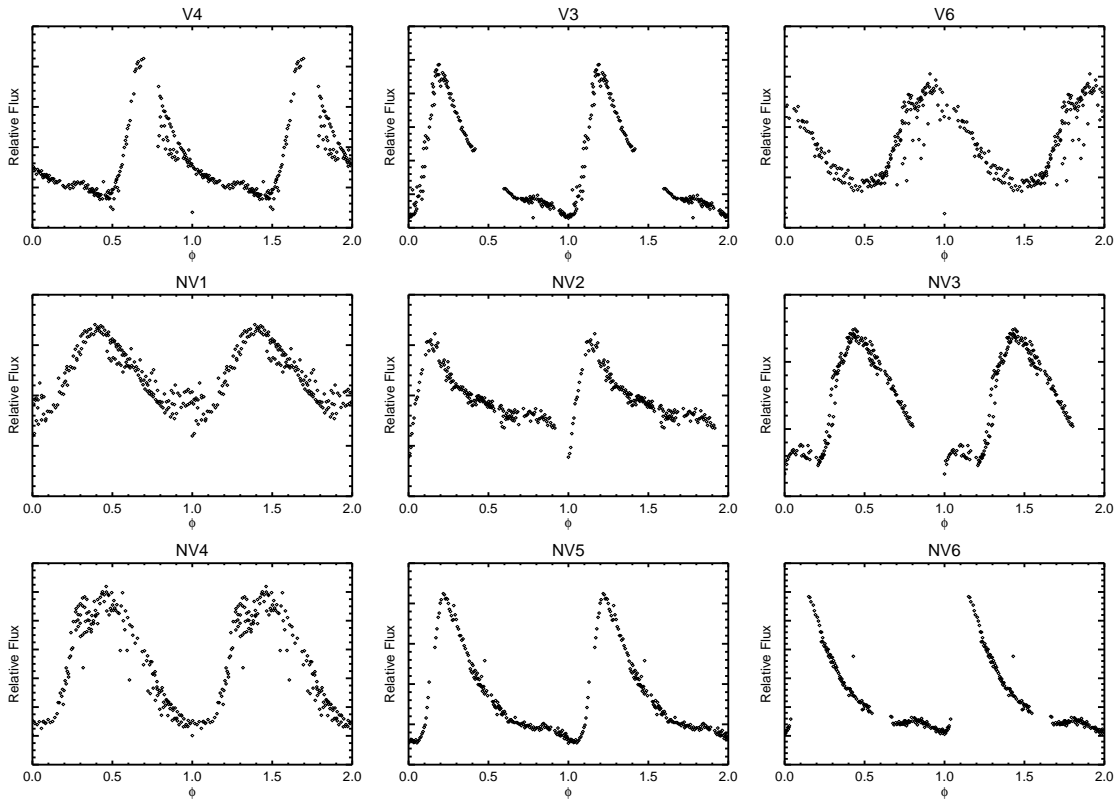}
  \caption{Light curves of the M79 RR Lyrae stars found in this study. The light curves are given in relative fluxes in the $B$ band.}
   \label{VM79}
   \end{figure*}

M79 ($\alpha_{J2000}$=05$^h$ 24$^m$ 10$^s$, $\delta_{J2000}$= $-24^\circ$  31' 27'') is a Galactic globular cluster
with an extended blue horizontal branch morphology. It is one of the seven clusters which have been associated with the CMa overdensity 
(e.g., Forbes et al. 2004). The variable stars in this cluster have been poorly studied, with the most recent time-series analysis dating
back to Rosino (1952).  
According to the Clement et al. (2001) online catalog, there are 8 known or suspected variable stars in M79, most of them
in the outskirts of the cluster, and only 3 have well-determined periods. 
In this work, we used 174 images in the $B$ filter, where the amplitude of RR Lyrae is enhanced, taken with the 1.54m Danish telescope in 2001.
An image subtraction analysis was performed using ISIS v2.2 (Alard 2000), which is expected to unveil previously unknown variables,
especially in the cluster core (see, e.g., Contreras et al. 2010 for a recent example of the technique's impressive detection power,
even in very crowded fields). 


Table~\ref{TaM79} displays the properties for the RR Lyrae stars found in this study. In Figure~\ref{VM79} we show the light curves (in relative fluxes) of the variables found in this study. 
V3, V4, and V6 are previously known variables, listed as in Clement's catalogue. 
NV1 displays an odd light curve, that could correspond to a double period~-- but further analysis is needed to put this result in a firmer footing.

The average period for the ab-type RR Lyrae is  $\langle P_{\rm ab} \rangle \simeq 0.68$~d, which corresponds to an OoII type. 
A future study will include the calibrated amplitudes obtained with DAOPHOT~II/ALLFRAME (Stetson 1994).


 \section{New photometry for NGC 1851}

   \begin{figure*}
   \centering
  \includegraphics[width=12cm]{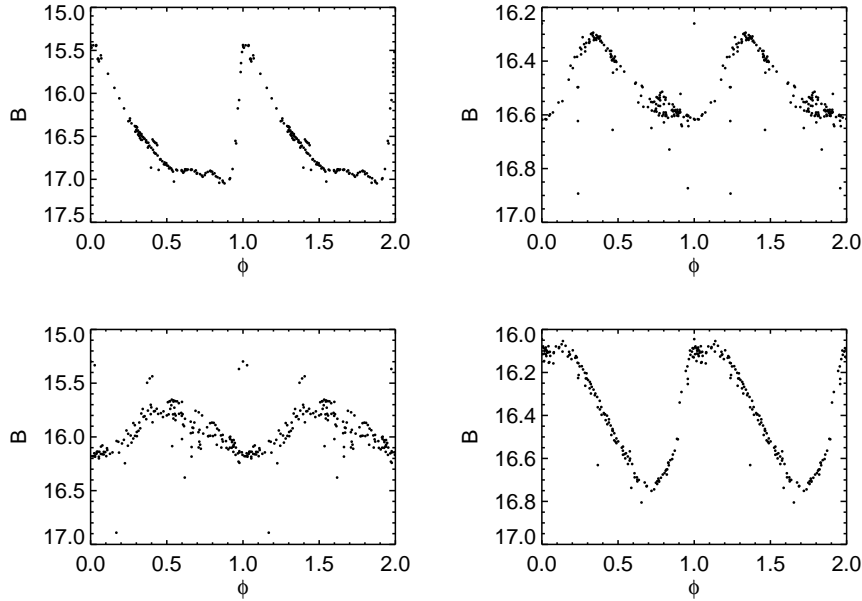}
  \caption{Light curves of the new RR Lyraes in NGC1851. }
   \label{VNGC1851}
   \end{figure*}
\begin{flushleft}
\begin{deluxetable*}{cccccl}[h!]
\tabletypesize{\normalsize}
\tablecaption{Details of four new RR Lyrae in NGC1851}
\tablewidth{0pt}
\tablehead{ \\ \colhead{Variable ID}   & \colhead{RA(J2000)} &
       \colhead{DEC(J2000)} & \colhead{Period (days)} & \colhead{Type} & \colhead{A$_B$} \\
}

\startdata
NV1 & 05 14 02.43 & -40 02 56.4 & 0.520589 & RRab & 1.60 \\
NV2 & 05 14 01.14 & -40 01 53.9 & 0.668266 & RRab?& 0.40 \\
NV3 & 05 14 05.08 & -40 02 40.2 & 0.279355 & RRc  & 0.55 \\
NV4 & 05 14 01.14 & -40 01 53.9 & 0.268519 & RRc  & 0.70  \\

 \enddata                                    
\label{TaN1851}
\end{deluxetable*}
\end{flushleft}

We present the result of a new photometric study of NGC~1851, another suggested member of the Canis Major system,  
based on more than 1000 images from archives, performed with DAOPHOT~II/ALLFRAME (Stetson 1994). 
We discovered 4 variables with reasonably well-measured periods, as shown in Figure~\ref{VNGC1851} (See Table~\ref{TaN1851}
for details).
We also recovered 29 of the previously known variables. 
For the RR Lyrae stars with reliable periods, we obtain in this study $\langle P_{ab} \rangle \simeq 0.569$~d, 
consistent with an OoI type, as found also in previous studies of the cluster (e.g., Walker 1998).

   \begin{figure*}
   \centering
  \includegraphics[width=15cm]{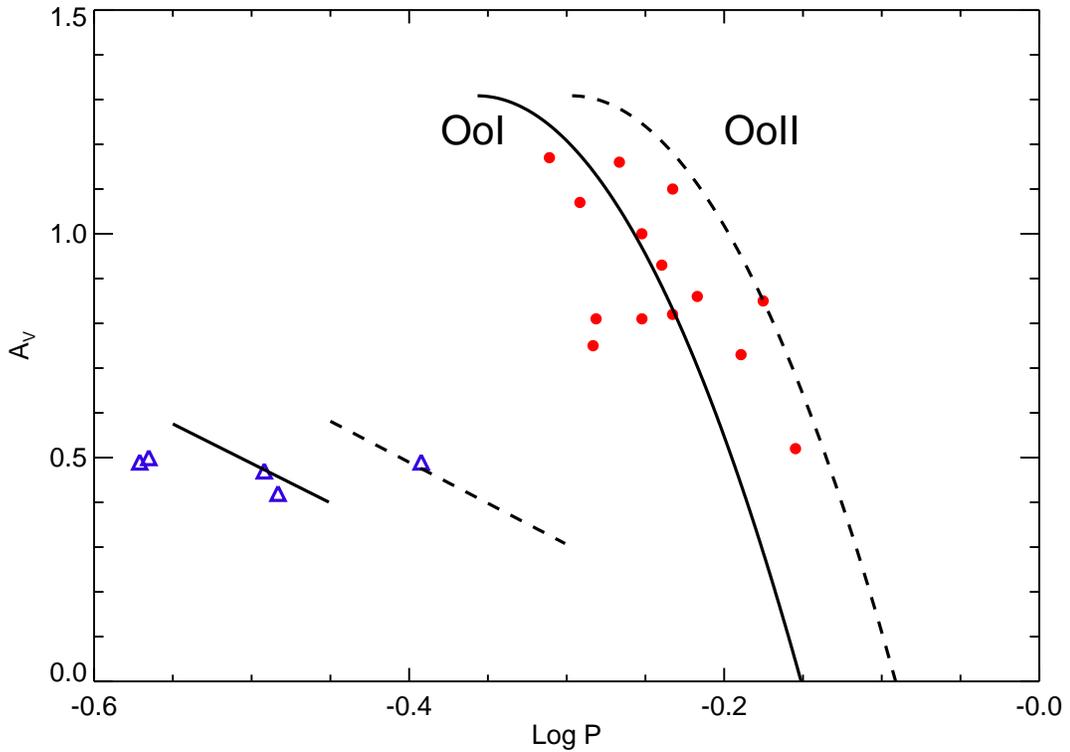}
  \caption{Period-amplitude diagram for the NGC~1851 RR Lyrae found in this study. Filled red dots correspond to ab-type 
     RR Lyrae, and empty blue triangles indicate RRc stars. Reference OoI and OoII lines are taken from Zorotovic et al. 
	 (2010), based on Cacciari et al. (2005).  }
   \label{Bailey}
   \end{figure*}

The period-amplitude diagram for NGC~1851 (Fig.~\ref{Bailey}) gives additional evidence on the cluster's Oosterhoff 
classification. We overplot reference lines corresponding to the different Oosterhoff types (from Zorotovic et al. 
2010). Again, we find that the RRab variables (red dots) fall predominantly on the locus defined by OoI  
clusters. RRc-type variables (blue triangles) follow a similar trend.

%
   \begin{figure*}
   \centering
   \includegraphics[width=10cm]{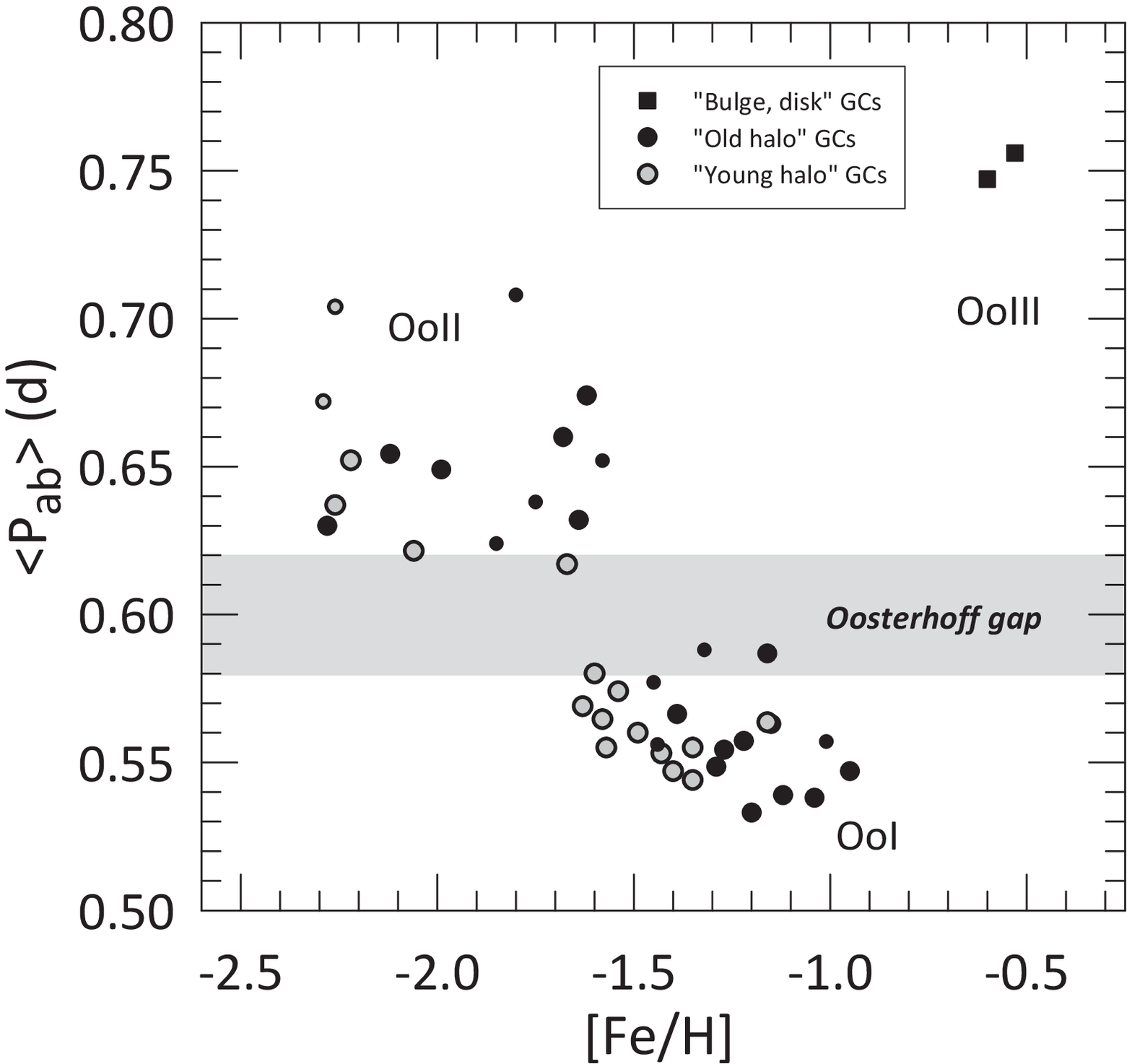}
   \includegraphics[width=10cm]{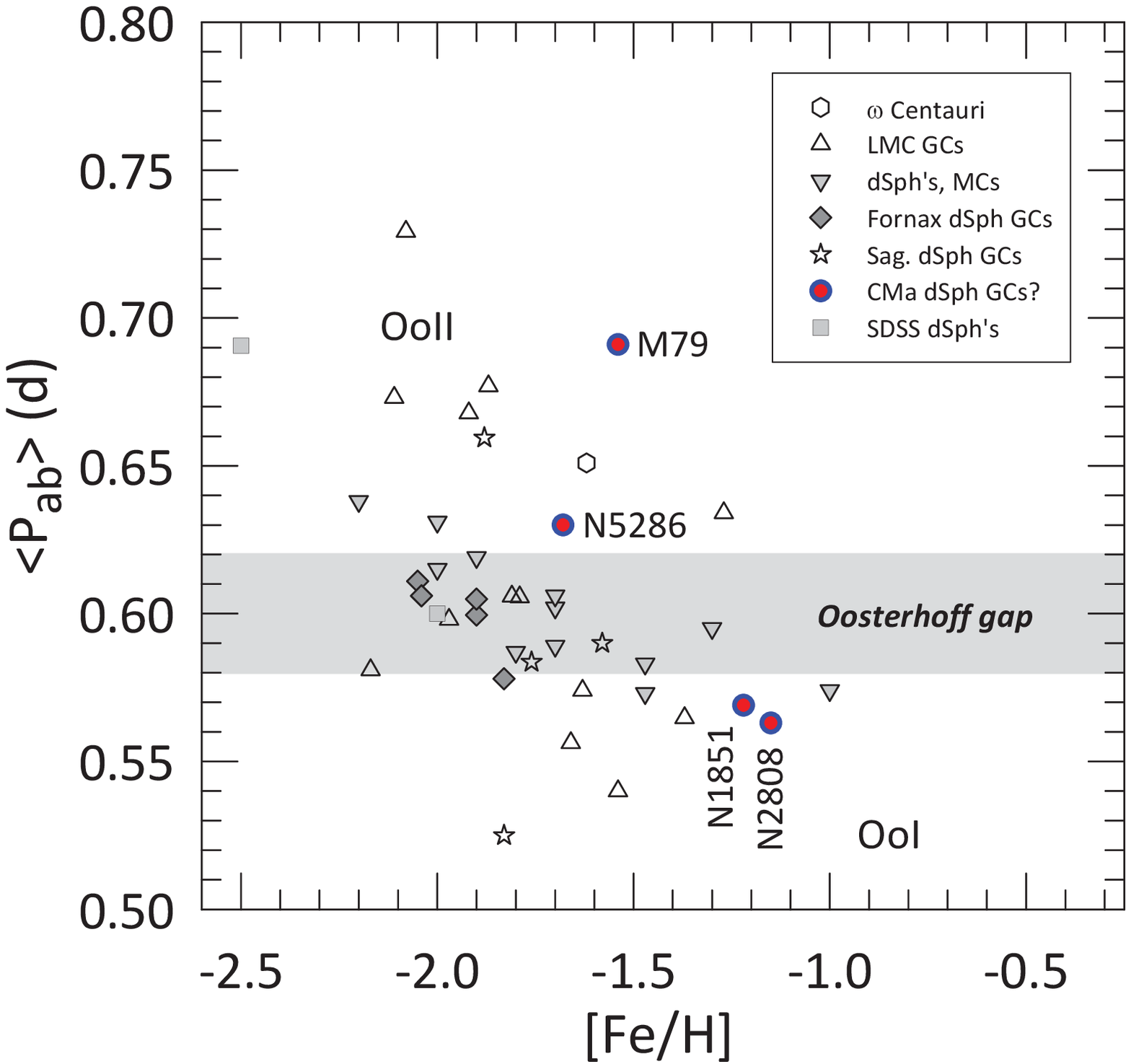}
  \caption{Mean ab-type RR Lyrae periods vs. metallicity diagram for Galactic ({\em upper panel}) and nearby extragalactic ({\em lower diagram})
    systems. In the lower panel, a few globular clusters that have previously been suggested to be associated with the CMa overdensity are 
	highlighted.}
   \label{oost}
   \end{figure*}
\section{On the Origin of the CMa Overdensity}
Through the analysis of the RR Lyrae population of these two clusters and the Oosterhoff argument, we can add some 
constraints that may be of relevance to the CMa debate. In particular, the inferred Oosterhoff types (OoII for M79, 
and OoI for NGC~1851) suggest that these systems follow the Galactic trend. This is clearly shown in Figure~\ref{oost}, 
adapted from Catelan (2009), where the average periods of ab-type RR Lyrae in Galactic (upper panel) and nearby 
extragalactic (lower panel) systems are shown. Our new measurements for M79 and NGC~1851 are also shown in the
lower panel of this figure, along with similar results for other CMa-related globular clusters, namely 
NGC~2808 (see Kunder et al. 2011, these proceedings) and NGC~5286 (Zorotovic et al. 2010). As can be seen, 
these four globular cluster conform to the Oosterhoff dichotomy. Even though this is not unprecedented for 
nearby extragalactic systems, it is certainly more common for systems of extragalactic origin to present 
Oosterhoff-intermediate types. This may represent additional evidence against an extragalactic origin for 
the CMa overdensity and its associated system of globular clusters. 

\vskip 1.025cm

\noindent {\bf Acknowledgments.} 
Support for P.A., M.C., and M.Z. is provided by the Chilean Ministry for the 
Economy, Development, and Tourism's Programa Inicativa Cient\'{i}fica Milenio 
through grant P07-021-F, awarded to The Milky Way Millennium Nucleus; the BASAL 
Center for Astrophysics and Associated Technologies (PFB-06); the FONDAP Center 
for Astrophysics (15010003); and Proyecto Fondecyt Regular \#1110326. P.A.
acknowledges additional support from SOCHIAS, PUC-DAA, and MECESUP. H.A.S thanks the US National Science Foundation for support. 
We also thank C. Contreras and A. Kunder for useful discussions. 


%

\end{document}